# Time–frequency distributions of Optical sources (II): externally modulated semiconductor lasers

**JOSÉ CAPMANY**[*]

*ITEAM Research Institute, Universitat Politècnica de València, Camino de Vera s/n, 46022 Valencia, Spain*

[*]*jcapmany@iteam.upv.es*

**Abstract:** We extend a time–frequency framework, previously developed for directly modulated lasers, to external modulation of the continuous-wave output of a 1550 nm DFB laser described by Langevin rate equations. Four modulators are considered: an InP electroabsorption modulator, a $LiNbO_3$ dual-drive Mach–Zehnder modulator, a dual-parallel Mach–Zehnder modulator and a silicon carrier-depletion modulator. For any memoryless modulator the chirp is shown to be purely transient, $\Delta\nu = \alpha_m(v)(4\pi)^{-1}\mathrm{d}\ln P/dt$, with a voltage-dependent chirp map $\alpha_m = \mathrm{Im}(M'/M)/\mathrm{Re}(M'/M)$ that the first moment of Cohen-class distributions recovers pointwise. Closed-form chirp maps, Jacobi–Anger line expansions, factorized Volterra kernels, intercept points and spurious-free dynamic ranges in 1 Hz are derived and validated. Because the modulators transfer the laser RIN unchanged, a low-power DFB source sets a common SFDR minimum near its relaxation frequency; compared with the intrinsic (upper-bound) DML model, the external links trade dynamic range for chirp control.

## 1. Introduction

External modulation of a continuous-wave (CW) semiconductor laser is the standard alternative to direct modulation when low chirp, high extinction, complex modulation formats or high analog linearity are required [1–3]. Four modulator families dominate practice. InP electroabsorption modulators (EAMs), based on the quantum-confined Stark effect [4,5], are compact and monolithically integrable with the laser, but exhibit a voltage-dependent chirp [6,7]. $LiNbO_3$ Mach–Zehnder modulators (MZMs) offer programmable chirp through dual-drive electrodes [1,6]. Dual-parallel (nested, IQ) MZMs synthesize arbitrary complex fields, including optical single-sideband (SSB) modulation, which is used to overcome dispersion-induced fading in radio-over-fiber links [8,9], and quadrature phase-shift keying (QPSK) [2]. Silicon carrier-depletion modulators, based on the plasma-dispersion effect [10,11], are the workhorse of silicon photonics, and their linearity has become a topic of interest for analog applications [12,13]. Thin-film $LiNbO_3$ is re-establishing the Pockels effect in integrated photonics [14].

In a companion paper (Part I) [15] we introduced a time–frequency description of directly modulated lasers (DMLs), in which the field predicted by the rate equations is represented by Cohen-class distributions [16,17]. Part I showed that for suitable kernels the first conditional moment of the distribution is exactly the laser chirp, and that the transient chirp is $\alpha$ times Cohen's instantaneous bandwidth. It then developed closed-form small-signal, line-expansion and two-tone (Volterra) analyses, including the relative intensity noise (RIN) from a Langevin treatment and the spurious-free dynamic range (SFDR) of analog links. Here we apply the same framework to external modulation of the CW output of the same laser, so that the two schemes can be compared on an equal footing.

The central observation is that, once its electrode response has been separated, an external modulator is a memoryless complex transfer $M(v)$ acting on the laser field. This has three consequences, which organize the paper. (i) The chirp has no adiabatic component. It is entirely determined by a voltage-dependent chirp map $\alpha_m(v)$, which generalizes the linewidth-enhancement factor and the chirp parameter of Koyama and Iga [6], and which the time–frequency distribution returns pointwise (Section 3). (ii) The spectra of sinusoidally modulated fields follow from Jacobi–Anger expansions of $M$, and their Wigner distributions from the line expansion of Part I (Section 4). (iii) The Volterra kernels of the link factorize into Taylor coefficients of the power transfer times the electrode responses. This gives closed-form intercept points, while the noise is dominated by the RIN of the laser, which the modulator transfers unchanged (Section 5). All closed-form results are validated numerically, and the four modulators are compared with the DML of Part I in digital (10 Gb/s non-return-to-zero and 10 GBd QPSK) and analog (one- and two-tone) operation.

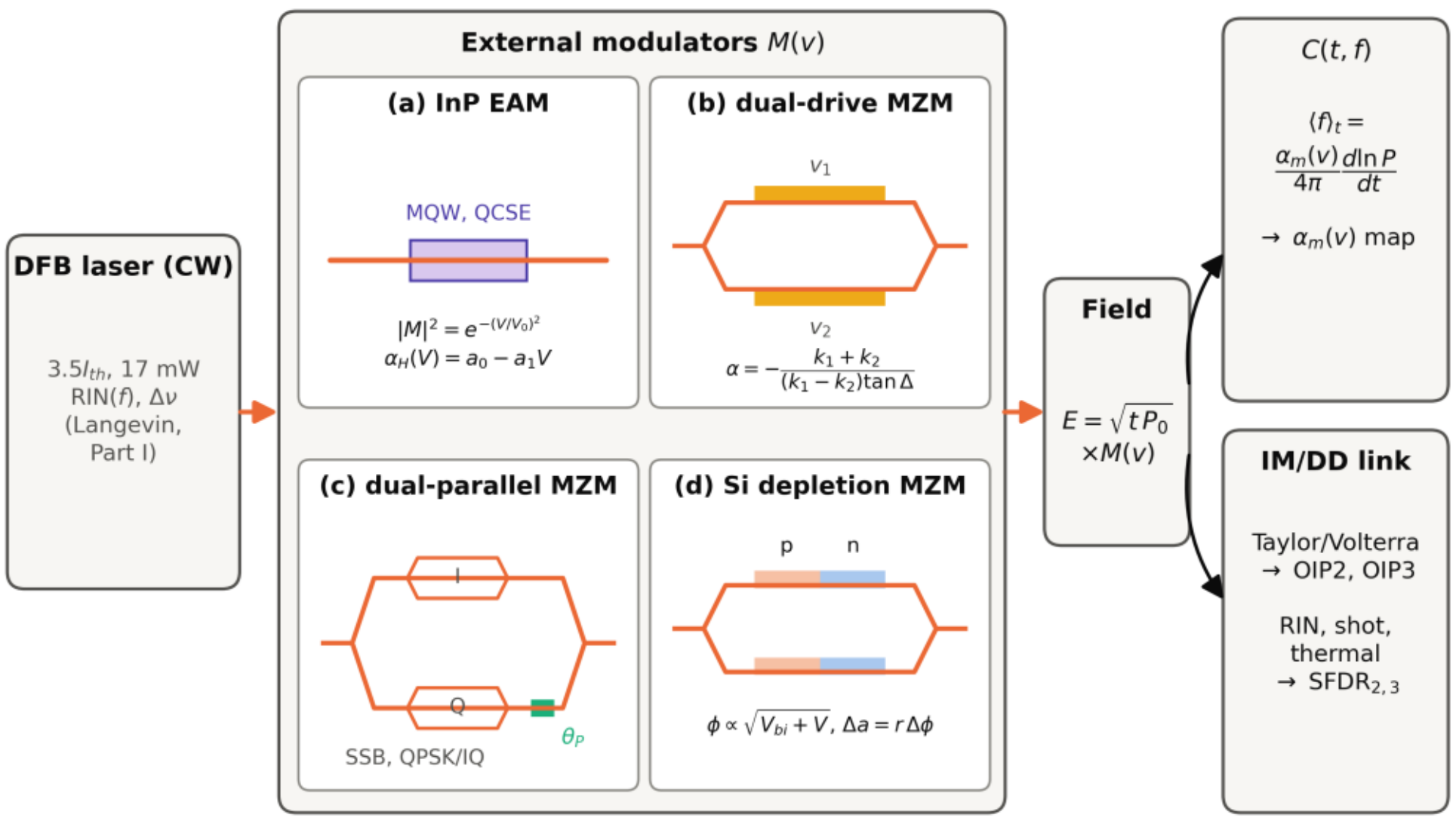

**Fig. 1.** Framework of this work. The CW output of the DFB laser of Part I (17.2 mW at $3.5I_{th}$, with the RIN and linewidth given by its Langevin rate equations) is modulated by (a) an InP electroabsorption modulator, (b) a dual-drive MZM, (c) a dual-parallel (IQ) MZM or (d) a silicon carrier-depletion MZM, each described by a memoryless field transfer $M(v)$ preceded by an electrode response. The output field is analyzed by Cohen-class distributions, which give the chirp map $\alpha_m(v)$, and in an intensity-modulation/direct-detection (IM/DD) link, where Taylor/Volterra kernels and the noise sources give the intercept points and the SFDRs.

## 2. Input field and modulator models

### *2.1 CW laser*

The optical carrier is the output of the DFB-1550 laser of Part I [15] biased at $3.5I_{th}$, which delivers $P_0 = 17.2$ mW with a relaxation frequency $f_R = 8.27$ GHz. Its RIN, obtained from the Langevin rate equations of Part I, is −164.0, −159.1 and −143.7 dB/Hz at 1, 2 and 10 GHz, with a peak at $f_R$. Its Lorentzian linewidth is 0.89 MHz, including the $\alpha$-enhancement. The phase diffusion accumulated over the 2 ns records used here has a variance of 0.011 rad², so the laser phase noise does not affect the time–frequency representations. Phase noise is also irrelevant for the back-to-back IM/DD links considered below. We therefore write the input field as $E_{in}(t) = \sqrt{P_0}$ in the deterministic analyses, and retain the laser RIN in the noise analysis.

### *2.2 General memoryless modulator*

Each modulator is modeled as a linear electrode, with transfer function $H_{el}(f)$ acting on the applied voltage $v(t)$, followed by a memoryless complex field transfer $M$:

$$E_{out}(t) = \sqrt{t_{ins}}\, E_{in}(t)\, M\big[v_f(t)\big], \qquad \tilde{v}_f(f) = H_{el}(f)\, \tilde{v}(f), \tag{1}$$

where $t_{ins}$ is the power transmission corresponding to the insertion loss and the tilde denotes the Fourier transform. For dual-input devices, $v$ is replaced by the pair $(v_I, v_Q)$. The output power and frequency deviation are

$$P_{out}(t) = t_{ins} P_0 \big|M(v_f)\big|^2, \qquad \Delta\nu(t) = \frac{1}{2\pi}\frac{d}{dt}\arg M\big[v_f(t)\big]. \tag{2}$$

Lumped electrodes (the EAM) are described by a single pole, $H_{el} = 1/(1 + if/f_{RC})$. Traveling-wave electrodes (the MZMs) are described by the standard velocity-mismatch and loss response for a matched, terminated line [1]

$$H_{el}(f) = \frac{1 - e^{-\Gamma(f)L}}{\Gamma(f)L}, \qquad \Gamma(f) = \alpha_{RF}\sqrt{f} + i\,\frac{2\pi f\,\Delta n}{c}, \tag{3}$$

where $L$ is the electrode length, $\alpha_{RF}$ the field attenuation coefficient of the line and $\Delta n$ the difference between microwave and optical group indices. The parameters are listed in Table 1; they are representative of each technology and do not describe specific commercial devices. For silicon, Eq. (3) is a simplification: the loss of pn-loaded lines is dominated by the junction series resistance and does not scale as $\sqrt{f}$, so the bandwidth given for the Si modulator is indicative.

### 2.3 InP electroabsorption modulator

The quantum-confined Stark effect shifts the excitonic absorption edge of the multiple-quantum-well core with the reverse bias $V$ [4,5]. We model the power transmission and the associated chirp as

$$|M(V)|^2 = \exp\left[-\left(\frac{V}{V_0}\right)^2\right], \qquad \alpha_H(V) = a_0 - a_1 V, \tag{4}$$

where $\alpha_H$ is the Henry factor of the EAM, which changes sign with bias as the operating point moves across the absorption edge [6,7]. Integrating $d\phi = (\alpha_H/2)\, d\ln|M|^2$ gives the phase in closed form:

$$M(V) = \exp\left[-\frac{V^2}{2V_0^2} - \frac{i}{V_0^2}\left(\frac{a_0 V^2}{2} - \frac{a_1 V^3}{3}\right)\right]. \tag{5}$$

With $V_0 = 1.40$ V the extinction is 19.9 dB at 3 V.

### 2.4 Dual-drive Mach–Zehnder modulator

A single drive voltage $v$ is applied to the two arms with weights $k_1$ and $k_2$ (push-pull, $k_1 = -k_2 = 1/2$; single drive, $k_1 = 1$, $k_2 = 0$), and a static phase difference $\theta_b$ sets the operating point:

$$M(v) = \frac{1}{2}\left[e^{i\theta_1(v)} + e^{i\theta_2(v)}\right] = \cos\Delta(v)\, e^{i\Sigma(v)}, \qquad \theta_{1,2} = \frac{\pi k_{1,2} v}{V_\pi} \pm \frac{\theta_b}{2}, \tag{6}$$

with $\Delta = (\theta_1 - \theta_2)/2$ and $\Sigma = (\theta_1 + \theta_2)/2$. Quadrature corresponds to $\theta_b = -\pi/2$.

### 2.5 Dual-parallel Mach–Zehnder modulator

Two push-pull child MZMs, driven by $v_I$ and $v_Q$ and biased at $b_I$ and $b_Q$, are combined with a parent phase $\theta_P$:

$$M(v_I, v_Q) = \frac{1}{2}\left[\cos\left(\frac{\pi v_I}{2V_\pi} + \frac{b_I}{2}\right) + e^{i\theta_P}\cos\left(\frac{\pi v_Q}{2V_\pi} + \frac{b_Q}{2}\right)\right]. \tag{7}$$

With $b_I = b_Q = \pi$ (children at null) and $\theta_P = \pi/2$, the children produce real fields between −1 and +1 and the device is an IQ modulator for QPSK. With $b_I = b_Q = -\pi/2$ (quadrature), $\theta_P = \pi/2$ and $v_Q$ equal to the Hilbert transform of $v_I$, it generates optical SSB with carrier [8,9].

### 2.6 Silicon carrier-depletion Mach–Zehnder modulator

Each arm contains a reverse-biased pn junction whose depletion width grows as $\sqrt{V_{bi} + V}$. The removal of free carriers raises the refractive index and lowers the absorption according to the Soref–Bennett relations [10,11]. We write the arm transmission as

$$A(V) = \exp[(r + i)\,\phi(V)], \qquad \phi(V) = K\left(\sqrt{V_{bi} + V} - \sqrt{V_{bi}}\right), \tag{8}$$

where $r$ is the ratio between the change of field attenuation and the change of phase. From the Soref–Bennett coefficients, $r \approx 0.03$ for holes and $0.12$ for electrons at doping levels near $5 \times 10^{17}$ cm$^{-3}$, and we use $r = 0.05$ for a junction in which both carriers are depleted. Since the hole contribution to the index scales as the 0.8 power of the density, $r$ depends weakly on the depletion depth; we neglect this dependence, which would add a small voltage dependence to $r$. The arms are reverse-biased at $V_B$ and driven differentially:

$$M(v) = \frac{A(V_B + v/2)\, e^{i\theta_b/2} + A(V_B - v/2)\, e^{-i\theta_b/2}}{2|A(V_B)|}. \tag{9}$$

The small-signal half-wave voltage at the bias point is $V_{\pi,pp} = 2\pi\sqrt{V_{bi} + V_B}/K = 6.70$ V. The phase bias $\theta_b$ represents a thermo-optic heater.

**Table 1. Modulator parameters. The electrical 3-dB bandwidth is the frequency where $|H_{el}|^2 = 1/2$, i.e., where the detected RF power of the link drops by 3 dB. $\alpha_{RF}$ is a field attenuation.**

| Parameter | InP EAM | $LiNbO_3$ DD-MZM | $LiNbO_3$ DP-MZM | Si depletion MZM |
|---|---|---|---|---|
| transfer model | Eqs. (4), (5) | Eq. (6) | Eq. (7) | Eqs. (8), (9) |
| key parameters | $V_0 = 1.40$ V, $a_0 = 1.5$, $a_1 = 1.2$ V$^{-1}$ | $V_\pi = 3.5$ V | $V_\pi = 3.5$ V (children) | $K = 1.57$ rad V$^{-1/2}$, $V_{bi} = 0.8$ V, $V_B = 2$ V, $r = 0.05$ |
| operating point | $V_b = 1.0$ V | quadrature, $\theta_b = -\pi/2$ | see text | quadrature (heater) |
| electrode | lumped, $f_{RC} = 35$ GHz | TW, $L = 3$ cm, $\alpha_{RF} = 0.35$ dB cm$^{-1}$ GHz$^{-1/2}$, $\Delta n = 0.03$ | as DD-MZM | TW, $L = 3$ mm, $\alpha_{RF} = 2.6$ dB cm$^{-1}$ GHz$^{-1/2}$, $\Delta n = 0.5$ |
| electrical 3-dB bandwidth | 35.0 GHz | 33.6 GHz | 33.6 GHz | 41.6 GHz |
| insertion loss $t_{ins}$ | 3 dB | 4 dB | 5 dB | 7 dB |

## 3. Chirp maps and time–frequency representation

### 3.1 The chirp map of a memoryless modulator

For a single-input modulator, write $M' = dM/dv$. Since $\ln M = \frac{1}{2}\ln|M|^2 + i\arg M$, Eq. (2) gives $d\ln P_{out}/dt = 2\,\mathrm{Re}(M'/M)\,\dot{v}_f$ and $2\pi\Delta\nu = \mathrm{Im}(M'/M)\,\dot{v}_f$. Eliminating $\dot{v}_f$,

$$\Delta\nu(t) = \frac{\alpha_m\left(v_f(t)\right)}{4\pi}\frac{d\ln P_{out}}{dt}, \qquad \alpha_m(v) = \frac{\mathrm{Im}[M'(v)/M(v)]}{\mathrm{Re}[M'(v)/M(v)]}. \tag{10}$$

Equation (10) is exact for any memoryless modulator and any waveform. Compared with the chirp of a DML [Eq. (6) of Part I], it contains no adiabatic term proportional to $P$, because the modulator has no memory of the optical power. The only difference with the transient chirp of a laser is that the coefficient $\alpha_m$ depends on the instantaneous voltage. We call $\alpha_m(v)$ the chirp map of the modulator. At a fixed operating point it reduces to the small-signal chirp parameter of Koyama and Iga [6].

The chirp maps of the four models follow directly from Eqs. (4)–(9). For the EAM, by construction, $\alpha_m(V) = \alpha_H(V) = a_0 - a_1 V$. For the DD-MZM, $M'/M = i\Sigma' - \Delta'\tan\Delta$, so

$$\alpha_m(v) = -\frac{k_1+k_2}{(k_1-k_2)\tan\Delta(v)}, \tag{11}$$

which vanishes for push-pull drive and equals ±1 at quadrature for single drive. For single drive, $\alpha_m$ vanishes at the transmission null and diverges at the transmission maximum, where $d\ln P/dv = 0$; the frequency deviation itself, $\Delta\nu = \Sigma'\dot{v}_f/2\pi$, remains finite and is proportional to the drive slew rate. For the Si modulator at quadrature ($v = 0$), the arms have equal and opposite derivatives, $A_{1,2}{}' = \pm\frac{1}{2}(r+i)\phi' A$, and $M'/M = (ir-1)\,\phi'\tan(\theta_b/2)/2$, so that

$$\alpha_m(0) = -r. \tag{12}$$

The same result, $\alpha_m(0) = -r$, holds at $v = 0$ for any heater bias, since $\theta_b$ cancels in the ratio. The residual chirp of a push-pull silicon modulator at its bias point is therefore exactly minus the Soref–Bennett amplitude-to-phase ratio of the depletion effect. Away from quadrature, the square-root dependence of $\phi$ makes the chirp map asymmetric. The DP-MZM has two inputs and no scalar chirp map. Its instantaneous frequency is $\Delta\nu = \frac{1}{2\pi}\frac{d}{dt}\arg M(v_I, v_Q)$, which becomes large whenever the field trajectory passes close to the origin of the complex plane.

### *3.2 Time–frequency representation*

The first conditional moment of any Cohen-class distribution whose kernel satisfies $\Phi(\nu,0) = 1$ and $\partial_\tau\Phi(\nu,0) = 0$ equals the instantaneous frequency (Part I). Using Cohen's instantaneous bandwidth $B_s = \frac{1}{4\pi}\,d\ln P/dt$ (with sign), Eq. (10) becomes

$$\langle f\rangle_t = \alpha_m\left(v_f(t)\right)B_s(t), \qquad \hat{\alpha}_m\left(v_f(t)\right) = \frac{\langle f\rangle_t}{B_s(t)}. \tag{13}$$

The ratio between the centroid and the instantaneous bandwidth of the distribution therefore returns the chirp map point by point, wherever the power changes. For a DML, by contrast, the centroid also contains the adiabatic term, so that $\langle f\rangle_t$ is not a single-valued function of $B_s$. In the Koch–Linke regression of Part I the adiabatic coefficient of an external modulator must vanish.

## 4. Sinusoidal modulation: line expansions

For a single-tone drive $v = V_m\cos x$, $x = 2\pi f_m t$, the modulated field is periodic and $E_{out} \propto \sum_n c_n\, e^{inx}$. For the DD-MZM, the Jacobi–Anger expansion of each arm gives, exactly,

$$c_n = \frac{i^n}{2}\left[J_n(m_1)\,e^{i\theta_{10}} + J_n(m_2)\,e^{i\theta_{20}}\right], \qquad m_{1,2} = \frac{\pi k_{1,2}V_m}{V_\pi}, \tag{14}$$

where $\theta_{10}$ and $\theta_{20}$ are the static arm phases. For the DP-MZM driven with $v_I = V_m\cos x$ and $v_Q = V_m\sin x$, writing $m = \pi V_m/(2V_\pi)$ and expanding each child,

$$c_n = \frac{1}{4}J_n(m)\left[i^n e^{ib_I/2} + (-i)^n e^{-ib_I/2} + e^{i\theta_P - in\pi/2}\left(i^n e^{ib_Q/2} + (-i)^n e^{-ib_Q/2}\right)\right]. \tag{15}$$

For $b_I = b_Q = -\pi/2$ and $\theta_P = \pi/2$, the $n = -1$ term cancels identically; this is SSB with carrier [8]. The residual sidebands at $n = -2$ and $n = +2$, together with the carrier and the $n = +1$ sideband, fix the purity of the SSB signal. For the EAM and the Si modulator, the $c_n$ follow from the Fourier series of $M(V_b + V_m\cos x)$; to lowest order $c_{\pm1} \simeq M'(V_b)V_m/2$ and $c_{\pm2} \simeq M''(V_b)V_m^2/8$. Since $M(V_b + V_m\cos x)$ is even in $x$, every memoryless single-input modulator driven by a sinusoid has $c_{-n} = c_n$: the line powers are exactly symmetric, and the chirp appears only in the phases of the lines, relative to those of an unchirped modulator. The pseudo-Wigner distribution of the modulated field is then given by the line expansion of Part I [Eq. (24) there], with these $c_n$: lines at half-integer multiples of $f_m$, with the auto-terms on the integer lines and interference terms on the half-integer lines and superposed on the integer lines.

## 5. Two-tone modulation, noise and dynamic range

### 5.1 Factorized Volterra kernels and intercept points

Let $p(v) = |M(v)|^2$ be the normalized power transfer at the operating point, with Taylor coefficients $p_k = d^k p/dv^k$. Because the modulator is memoryless after the electrode, the Volterra kernels of the optical power factorize:

$$H_n(\omega_1, \dots, \omega_n) = \frac{p_n}{n!} \prod_{j=1}^{n} H_{el}\left(\omega_j\right). \quad (16)$$

This is in contrast to the DML of Part I, whose kernels contain the dynamics of the laser. For the two-tone drive $v = V(\cos\omega_1 t + \cos\omega_2 t)$, the amplitudes of the power components relative to $t_{ins}P_0$ at $f_1$, $f_1 + f_2$ and $2f_1 - f_2$ are

$$\hat{p}_{f_1} = |p_1 H_1| V, \qquad \hat{p}_{f_1+f_2} = \frac{|p_2 H_1 H_2|}{2} V^2, \qquad \hat{p}_{2f_1-f_2} = \frac{|p_3 H_1^2 H_2|}{8} V^3, \quad (17)$$

with $H_j = H_{el}\left(\omega_j\right)$; the phase of the IMD3 term involves $H_1^2 H_2^*$, but only magnitudes enter below. The input voltage intercepts per tone are therefore

$$V_{IIP2} = \frac{2|p_1|}{|p_2 H_2|}, \qquad V_{IIP3}^2 = \frac{8|p_1|}{|p_3 H_1 H_2|}. \quad (18)$$

For the push-pull MZM at quadrature, $p(v) = \frac{1}{2}[1 + \sin(\pi v/V_\pi)]$, so $p_2 = 0$ (no second-order distortion) and $V_{IIP3}^2 = 8V_\pi^2/(\pi^2|H_1 H_2|)$. This gives the classical input intercept IIP3 $= 4V_\pi^2/(\pi^2 R_{in}|H_1 H_2|)$ [3,18]. For the EAM, with $u = V/V_0$ and $T = e^{-u^2}$,

$$p_2 = \frac{(4u^2-2)T}{V_0^2}, \qquad p_3 = \frac{(12u-8u^3)T}{V_0^3}, \quad (19)$$

so that the second-order distortion vanishes at $V = V_0/\sqrt{2} = 0.99$ V and the third-order distortion at $V = \sqrt{3/2}\,V_0 = 1.715$ V. Linearization bias points of this kind are exploited in EAM links; here they follow in closed form for the exponential-quadratic model, and their exact values depend on the absorption law of the device. For the Si modulator with $r = 0$, the differential phase $\phi(V_B + v/2) - \phi(V_B - v/2)$ is odd in $v$ despite the square-root law, and $p_2$ vanishes at quadrature. The second-order distortion at quadrature therefore originates in the absorption modulation: $p_2 = -7.7 \times 10^{-4}$ V$^{-2}$ for $r = 0.05$, against $p_1 = 0.235$ V$^{-1}$. It can be cancelled by a small offset of the heater bias. For the SSB DP-MZM with a 90° hybrid, the input power is shared by the two children. Since $p\left(v_I, v_Q\right) = \frac{1}{4} + \frac{1}{8}\left[\sin(\pi v_I/V_\pi) + \sin\left(\pi v_Q/V_\pi\right)\right]$ at the SSB operating point, the detected intensity is the sum of the two child intensity modulations. It is separable in $v_I$ and $v_Q$ at $\theta_P = \pi/2$, so no second-order intermodulation appears for any drive level; a 1° error in $\theta_P$ or in a child bias limits SFDR$_2$ to 94 dB·Hz$^{1/2}$.

### 5.2 Noise and SFDR

The link is the IM/DD link of Part I: a photodiode of responsivity $\mathcal{R}$ delivers the photocurrent $i = \mathcal{R}P_{out}$ into $R_L = 50\ \Omega$, and the modulator input is matched to $R_{in} = 50\ \Omega$, so the input power per tone is $V^2/(2R_{in})$. The noise density is $k_B T + 2qI_{dc}R_L + \mathrm{RIN}(f) I_{dc}^2 R_L$, with $I_{dc} = \mathcal{R} t_{ins} P_0 \overline{p}$. Here $\overline{p}$ is the mean normalized transmission and RIN is that of the CW laser: a memoryless modulator multiplies the power by a deterministic factor and transfers the relative intensity fluctuations unchanged. The SFDRs in 1 Hz follow from Eq. (17) exactly as in Part I, SFDR$_2 = \frac{1}{2}[\mathrm{OIP2} - \mathcal{N}(f_1 + f_2)]$ and SFDR$_3 = \frac{2}{3}[\mathrm{OIP3} - \mathcal{N}(2f_1 - f_2)]$. Near a distortion null the small-signal intercepts diverge. We therefore also compute the SFDR from the complete harmonic balance, as the ratio of the fundamental to the noise at the drive level where the distortion product equals the noise in 1 Hz. This quantity includes all higher-order contributions.

## 6. Numerical method

The drive waveforms are those of Part I: 10 Gb/s non-return-to-zero (NRZ) sequences of rectangular pulses filtered by a first-order RC network ($\tau_{RC} = 15$ ps), sinusoids and two tones. The modulator electrode response is applied in the frequency domain with zero padding, and the field is sampled at 1 ps, like the laser records of Part I. For the QPSK example the I and Q children are driven by two independent RC-filtered bit sequences at 10 GBd. Harmonic balance for the two-tone analysis is performed exactly for the memoryless models by sampling the power transfer on a 64 × 64 grid of the two tone phases, after applying $H_{el}(f_1)$ and $H_{el}(f_2)$, and taking a two-dimensional FFT. The resulting intermodulation amplitudes are exact to machine precision for all drive levels up to aliasing of orders above 31. The CW laser RIN is taken from Eq. (37) of Part I at $3.5I_{th}$. Table 2 summarizes the operating conditions.

**Table 2. Drive conditions (10 Gb/s NRZ with $\tau_{RC} = 15$ ps unless stated otherwise).**

| Modulator | Digital drive | Analog operating point |
|---|---|---|
| InP EAM | $V$ between 0.4 V (“1”) and 2.4 V (“0”) | $V_b = 1.0$ V (swept 0.4–2.4 V) |
| DD-MZM, push-pull | $v = \pm V_\pi/2$ about quadrature | quadrature (bias error swept ±20°) |
| DD-MZM, single drive | $v = \pm V_\pi/2$ about quadrature | – |
| Si depletion MZM | $v = \pm 2$ V (4 $V_{pp}$) about quadrature | quadrature (heater bias swept ±30°) |
| DP-MZM | QPSK, $v_{I,Q} = \pm V_\pi$, children at null, $\theta_P = \pi/2$ | SSB: children at quadrature, $\theta_P = \pi/2$, 90° hybrid |
| DML (Part I) | $1.2I_{th} \rightarrow 3.5I_{th}$ | $I_b = 3.5I_{th}$ |

## 7. Results

### *7.1 Transfer functions and chirp maps*

Figure 2 shows the static power transfers and the chirp maps. The EAM chirp map computed numerically from $M(V)$ coincides with $\alpha_H(V)$ and changes sign at $V = 1.25$ V; the IMD2 and IMD3 null biases of Eq. (19) are marked on the transfer. The three DD-MZM drive configurations have identical power transfers. Their chirp maps, however, are 0 (push-pull), 1 (single drive) and 0.4 ($k_1 = 0.7$, $k_2 = -0.3$) at quadrature, in agreement with Eq. (11) (markers). For asymmetric drive, the chirp map vanishes at the null and diverges at the transmission maximum, where the power is stationary. For the Si modulator, the chirp map equals −0.050 at quadrature, as predicted by Eq. (12). It is asymmetric in the drive voltage because of the square-root phase response: −0.073 at $v = -2$ V and −0.56 at $v = +2$ V. For $r = 0$, the chirp at quadrature vanishes but the asymmetry remains.

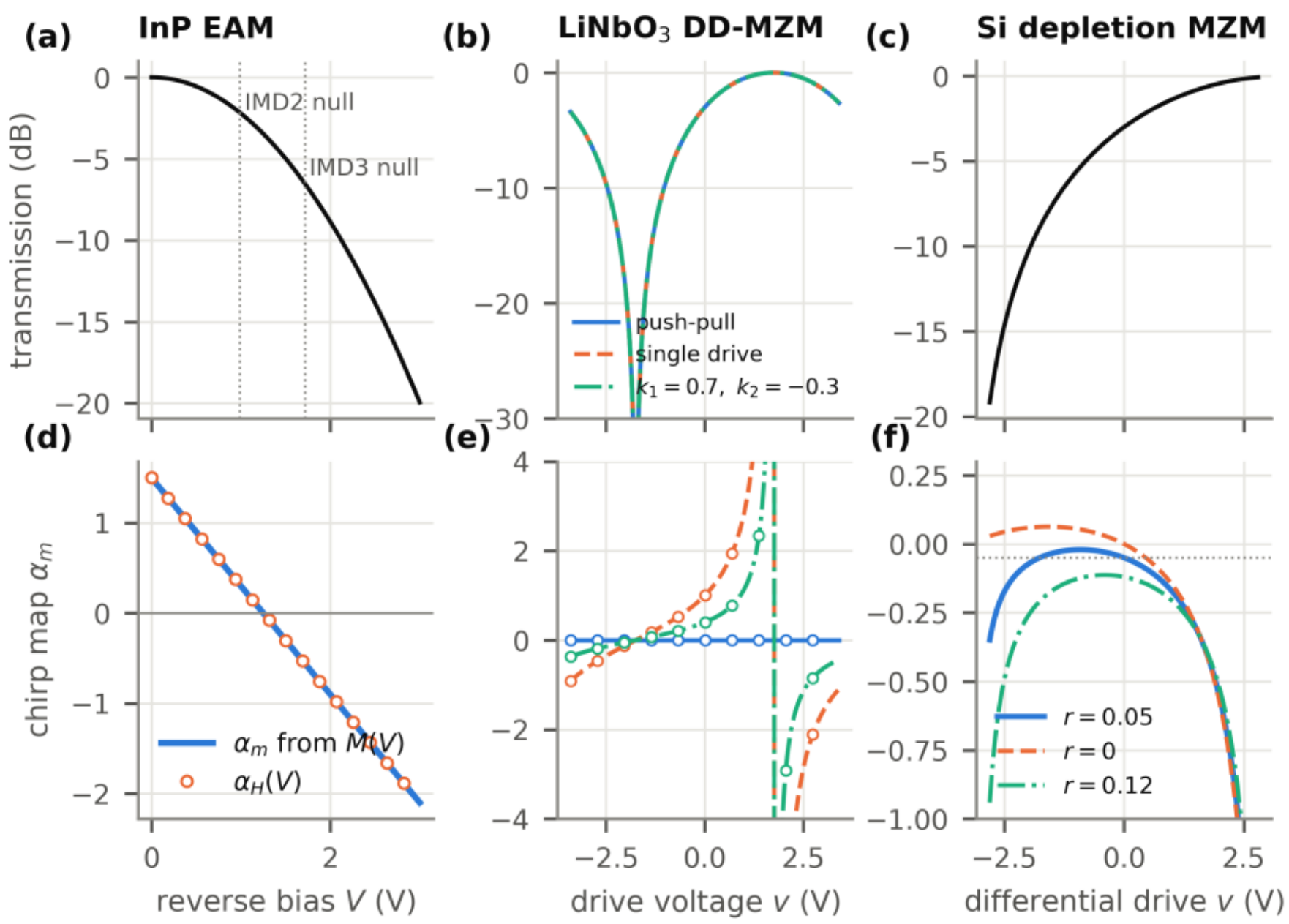

**Fig. 2.** Static power transfer (top) and chirp map (bottom). (a), (d) InP EAM: transmission versus reverse bias, with the IMD2 and IMD3 null biases of Eq. (19), and $\alpha_m$ from $M(V)$ (line) compared with $\alpha_H(V)$ (circles). (b), (e) DD-MZM for push-pull (solid), single-drive (dashed) and asymmetric ($k_1 = 0.7$, $k_2 = -0.3$, dash-dotted) configurations, with Eq. (11) as circles. (c), (f) Si depletion MZM at quadrature for $r = 0.05$ (solid), $r = 0$ (dashed) and $r = 0.12$ (dash-dotted); the dotted line marks $-r$ for $r = 0.05$.

### *7.2 Digital modulation*

Figure 3 compares the pseudo-Wigner distributions (PWVD, 40 ps lag window) of 10 Gb/s NRZ signals generated by the DML of Part I and by the external modulators, together with a 10 GBd QPSK signal from the DP-MZM. The chirp statistics are listed in Table 3. The DML chirp spans 55.8 GHz peak to peak, with the relaxation-oscillation strokes and adiabatic offsets discussed in Part I. The EAM chirp consists only of short transient excursions at the transitions, whose sign follows $\alpha_m(V)$: the rising edge, which starts in the high-bias, negative-$\alpha$ region, is red-shifted, and the falling edge shows a small red and then blue excursion as the bias crosses the zero of the chirp map. There is no frequency offset between the "1" and "0" levels. The push-pull MZM produces a distribution symmetric about the carrier, with zero chirp and 38.5 dB extinction. The single-drive MZM chirp is proportional to the drive slew rate, $\Delta\nu = \dot{v}_f / 4V_\pi$, independently of the operating point, and appears as bounded strokes of ±10 GHz at every transition. The Si modulator has a residual chirp of only 1.2 GHz peak to peak, at a lower extinction of 9.7 dB. In the QPSK signal from the DP-MZM the power dips at every symbol transition; simultaneous transitions of I and Q take the field through the origin, where the phase jumps by π, and the instantaneous frequency shows spikes of up to −25 GHz (ideally a delta function, whose sampled height depends on the 1 ps grid and the power mask), while the distribution shows interference stripes between consecutive symbols.

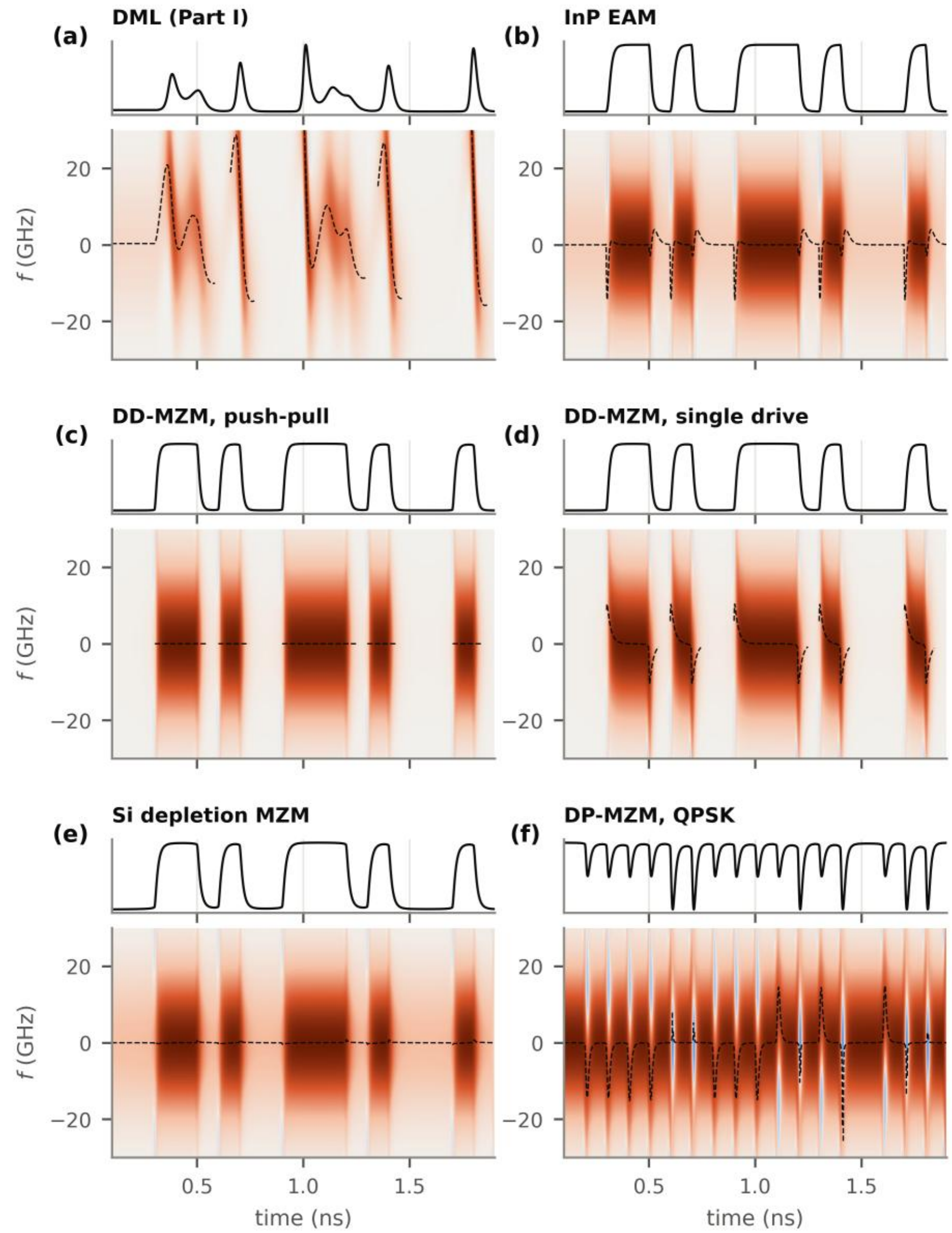

**Fig. 3.** Output power (top traces, arbitrary scale) and PWVD ($T_h = 40$ ps) with the instantaneous frequency (dashed, shown where $P$ exceeds 2% of the peak) for (a) the DML of Part I, (b) the InP EAM, (c) the push-pull and (d) the single-drive DD-MZM and (e) the Si depletion MZM, all with the same 10 Gb/s RC-filtered NRZ sequence, and (f) a 10 GBd QPSK signal from the DP-MZM.

**Table 3. Chirp and extinction of the digital signals of Fig. 3. The peak-to-peak chirp is taken over samples above 2% of the peak power and the rms chirp is power-weighted over samples above 5%. The extinction ratio is the ratio of the 99th to the 1st percentile of the power record, which includes the finite transition times.**

| Source | Chirp peak-to-peak [GHz] | rms chirp [GHz] | Extinction ratio [dB] |
|---|---|---|---|
| DML, 1.2 → 3.5$I_{th}$ | 55.8 | 12.19 | 36.4 |
| InP EAM, 2.4 → 0.4 V | 18.3 | 1.27 | 12.4 |
| DD-MZM, push-pull | 0 | 0 | 38.5 |

| Source | Chirp peak-to-peak [GHz] | rms chirp [GHz] | Extinction ratio [dB] |
|---|---|---|---|
| DD-MZM, single drive | 20.7 | 2.83 | 38.5 |
| Si depletion MZM, 4 $V_{pp}$ | 1.2 | 0.15 | 9.7 |
| DP-MZM, QPSK | 40.6 | 2.90 | – |

### *7.3 Chirp-map extraction from the distribution*

Figure 4 applies Eq. (13) to the NRZ records of Fig. 3, using the centroid of a PWVD with $T_h = 20$ ps at the samples where the power exceeds 5% of the peak and the instantaneous bandwidth exceeds 15% of its maximum. For the EAM the ratio $\langle f \rangle_t / B_s$, plotted against the instantaneous voltage, reproduces the chirp map over 0.5–2.4 V within 0.08 (345 samples). For the single-drive MZM it does so within 0.007 over −1.25 to −0.1 V, and for the Si modulator within 0.018 over −1.8 to +1.5 V. A single NRZ record thus measures the chirp map over the part of the voltage excursion where the power changes significantly. Figure 4(d) contrasts the DML and the EAM in the plane of the centroid versus the signed instantaneous bandwidth. The EAM points lie on the one-dimensional curve $(B_s, \alpha_m(v) B_s)$ parameterized by the voltage, whereas the DML points spread vertically because of the adiabatic term. The four-term Koch–Linke regression of Part I applied to the same records confirms this: it gives $\alpha = 3.001$ and $\alpha\kappa/4\pi = 0.2795$ GHz/mW for the DML, and $\alpha\kappa/4\pi = -1.3 \times 10^{-4}$ GHz/mW, i.e., no adiabatic chirp, for the EAM. For the EAM the fitted $\alpha = -1.28$ is only an average of the chirp map over the transitions.

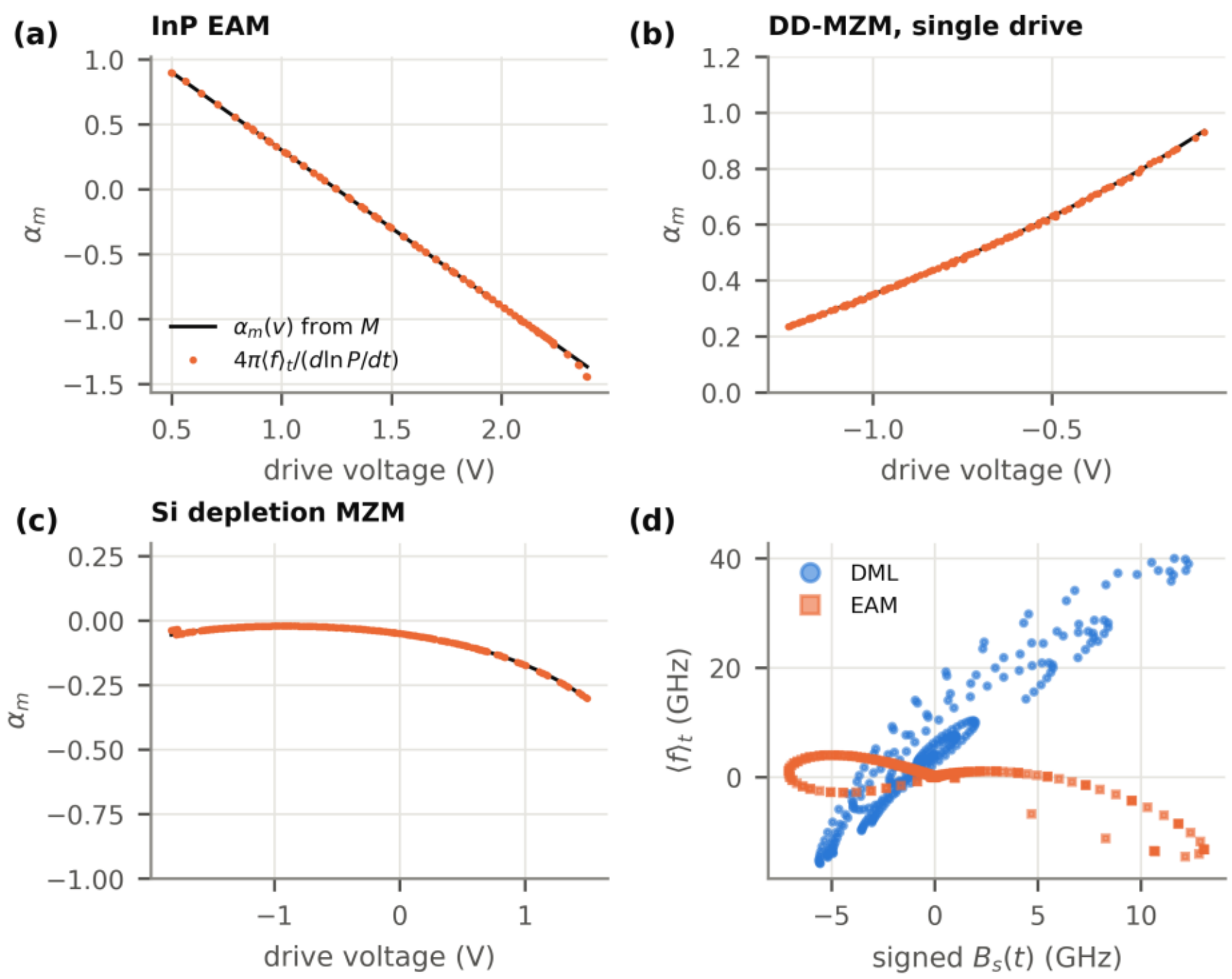

**Fig. 4.** Chirp map recovered from the distribution, $4\pi\langle f \rangle_t/(d\ln P/dt)$ (dots), versus the instantaneous drive voltage, compared with $\alpha_m(v)$ from the device model (line), for (a) the InP EAM, (b) the single-drive DD-MZM and (c) the Si depletion MZM. (d) Centroid of the PWVD versus the signed instantaneous bandwidth for the DML (circles) and the EAM (squares).

### *7.4 Single-tone modulation*

Figure 5 shows the line spectra for $f_m = 5$ GHz and $V_m = 1$ V. The closed forms, Eqs. (14) and (15), agree with the FFT of the modulated field to machine precision (largest deviation $6 \times 10^{-17}$ in $|c_n|$); this is a consistency check of the expansions rather than an independent validation. The push-pull MZM produces a symmetric double-sideband spectrum. The single-drive MZM produces a broader spectrum that is also exactly symmetric in power; its chirp changes only the phases of the lines, which is what causes dispersion-induced fading [9]. The DP-MZM suppresses the $n = -1$ sideband completely. Its strongest residual lines are at $n = \pm 2$, 31.7 dB below the carrier and 21.9 dB below the $n = +1$ sideband. The EAM spectrum is also exactly symmetric in power. The PWVDs with a 1 ns lag window [Figs. 5(e) and (f)] display the line structure predicted by the line expansion of Part I. In the double-sideband case, interference terms at ±2.5 GHz oscillate in sign at the modulation frequency. In the SSB case the terms at negative frequencies are strongly suppressed, and the remaining structure is concentrated on the positive-frequency side.

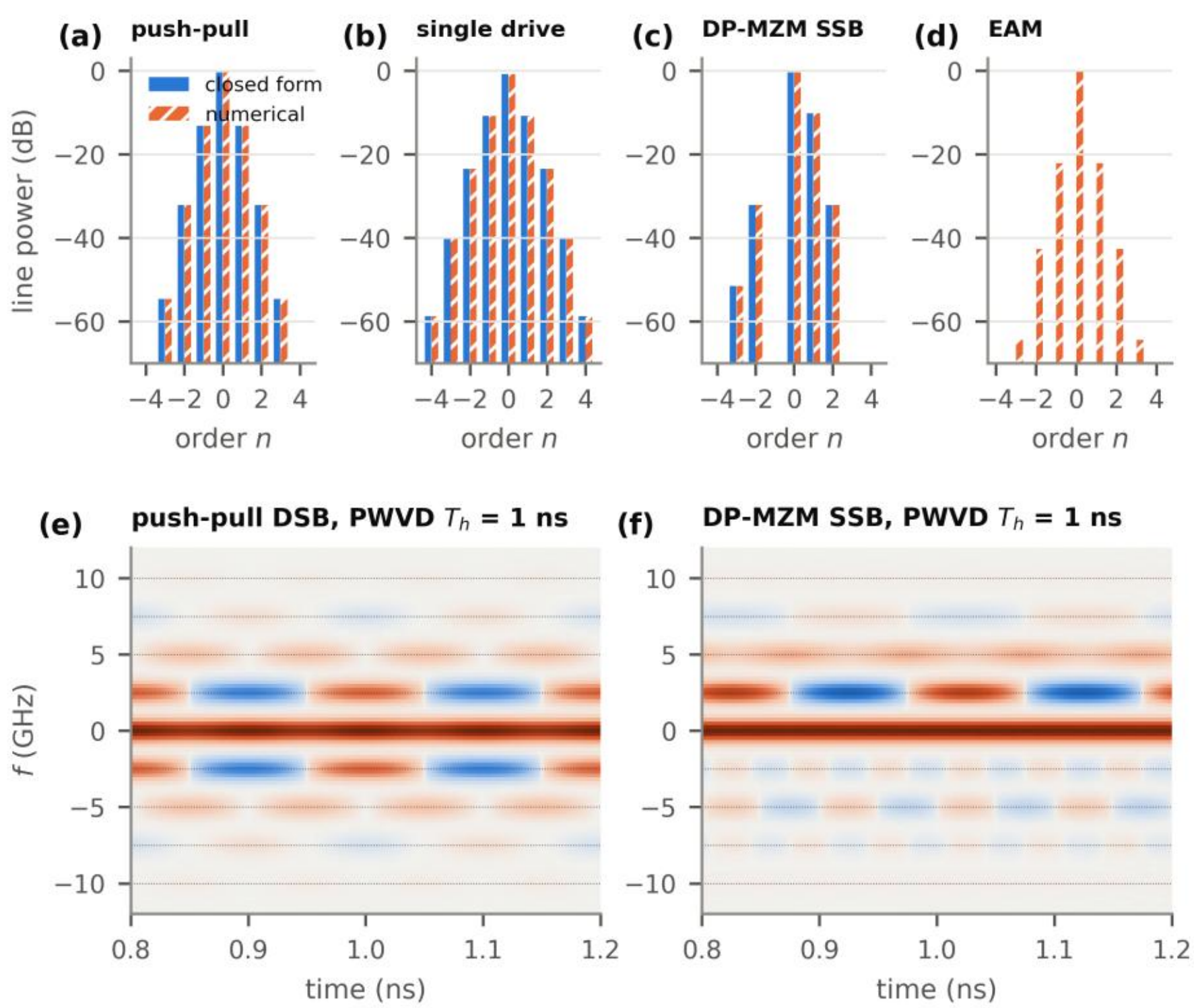

**Fig. 5.** Single-tone modulation at $f_m = 5$ GHz, $V_m = 1$ V. Line powers relative to the total power for (a) the push-pull and (b) the single-drive DD-MZM, (c) the DP-MZM in SSB configuration and (d) the EAM ($V_b = 1$ V, $V_m = 0.3$ V), from the closed forms, Eqs. (14) and (15) (solid bars; not shown for the EAM), and from the FFT of the field (hatched bars). PWVD with $T_h = 1$ ns for (e) the push-pull MZM and (f) the SSB DP-MZM; dotted lines mark multiples of $f_m/2$.

### *7.5 Two-tone intermodulation and SFDR*

Figure 6 shows the fundamental, IMD2 ($f_1 + f_2$) and IMD3 ($2f_1 - f_2$) output powers versus the input power per tone at $f_1 = 2$ GHz ($f_2 = f_1 + 50$ MHz), computed by exact harmonic balance, together with the small-signal asymptotes. As a consistency check, the closed forms of Eqs. (17) and (18) agree with the numerical intercepts: for the push-pull MZM the IIP3 is 20.70 dBm from the closed form and 20.71 dBm numerically, and the link gain is −18.94 dB in both cases. For the EAM the IIP3 is 15.99 dBm in both cases. The push-pull MZM and the SSB DP-MZM produce no IMD2 at their nominal bias points. The EAM at $V_b = 1$ V, close to its IMD2 null, shows an IMD2 characteristic with a slope-2 region at low drive and a cancellation notch near −4 dBm, where the fourth-order term cancels the residual second-order one. The Si modulator at quadrature shows IMD2 at all drive levels, produced by the absorption modulation ($r \neq 0$).

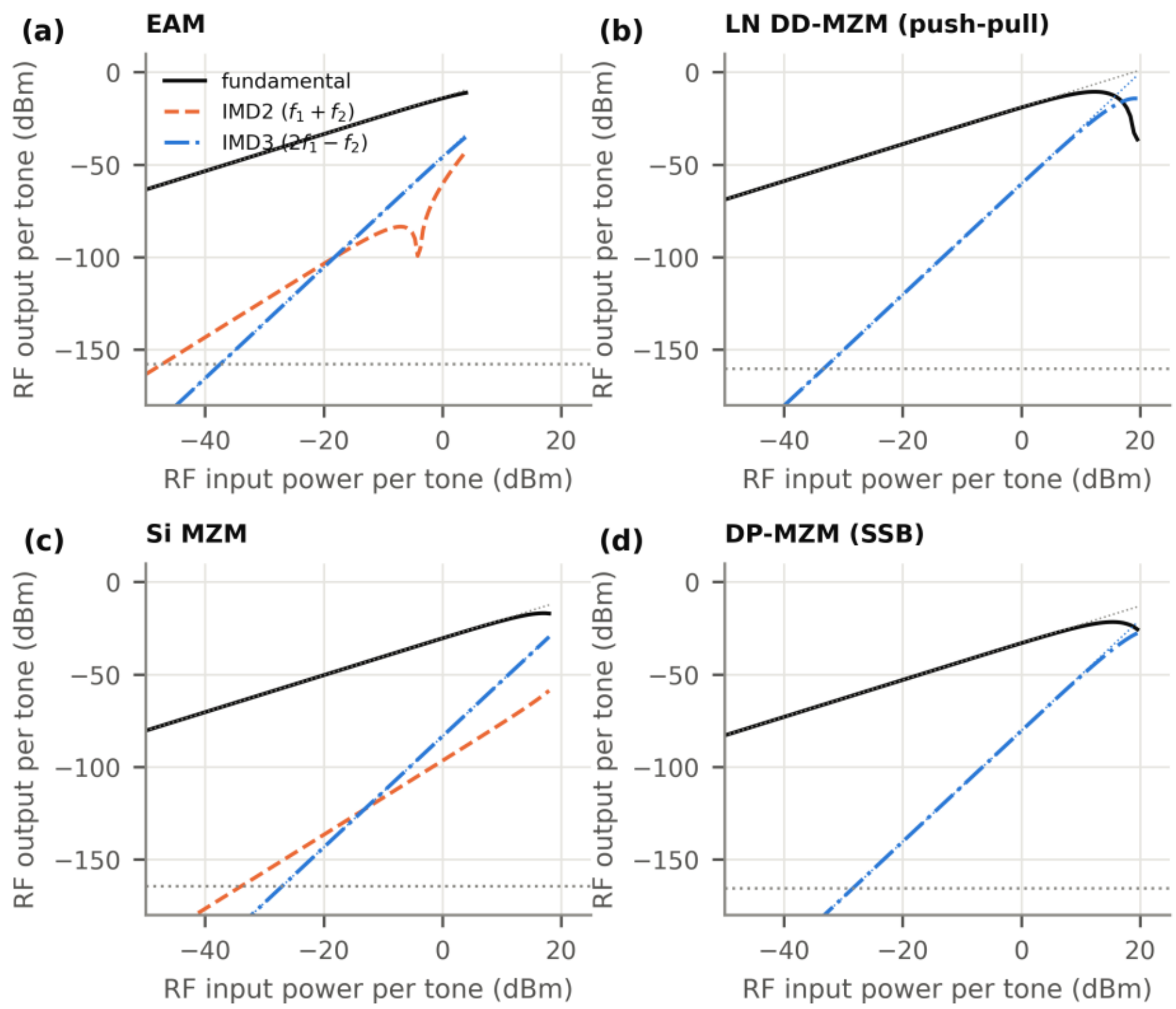

**Fig. 6.** RF output power per tone of the fundamental (solid), IMD2 at $f_1 + f_2$ (dashed) and IMD3 at $2f_1 - f_2$ (dash-dotted) versus RF input power per tone at $f_1 = 2$ GHz for (a) the EAM ($V_b = 1$ V), (b) the push-pull $LiNbO_3$ MZM, (c) the Si depletion MZM and (d) the DP-MZM in SSB configuration. Thin dotted lines: small-signal asymptotes of slope 1 and 3; horizontal dotted line: noise floor in 1 Hz at $2f_1 - f_2$. The EAM drive is limited to 0.5 V per tone so that the device remains reverse biased.

The link metrics are collected in Table 4. At 2 GHz, the four externally modulated links have $SFDR_3$ between 104.1 and 108.1 dB·Hz$^{2/3}$. The exact harmonic-balance SFDRs coincide with the small-signal values to 0.01 dB at the nominal bias points. The noise floors, between −157.7 and −165.5 dBm/Hz, are set by the laser RIN and the shot noise of the low photocurrents (1.1–4.1 mA). These photocurrents result from the insertion losses and the quadrature bias. Increasing the laser power to 100 mW at fixed RIN improves $SFDR_3$ by only 1.2–3.0 dB (to 107.1–109.7 dB·Hz$^{2/3}$), because the links then become RIN-limited. A laser with lower RIN would improve these values further. The DML link of Part I at the same laser bias reaches 115.6 dB·Hz$^{2/3}$ at 2 GHz, and still 114.0 dB·Hz$^{2/3}$ when attenuated to the same photocurrent as the MZM link (2.74 mA). The difference therefore lies in the intercept point rather than in the photocurrent. This advantage is not physical: the DML model contains no static nonlinearity of the light–current characteristic, no leakage or thermal effects and no parasitics, so its intercept is an intrinsic upper bound. The comparison should be read as a bound, not as a ranking of technologies.

**Table 4. Two-tone link metrics ($f_2 = f_1 + 50$ MHz, $\mathcal{R} = 0.8$ A/W, $R_L = R_{in} = 50\ \Omega$, laser at $3.5I_{th}$, 17.2 mW). "—": no IMD2 at the ideal operating point. With a 1° bias error, $SFDR_2$ is 94.6–94.7 dB·Hz$^{1/2}$ for the MZM and 94.0 dB·Hz$^{1/2}$ for the DP-MZM (1° error of either the parent phase or a child bias).**

| Link | $f_1$ [GHz] | Gain [dB] | $I_{dc}$ [mA] | OIP2 [dBm] | OIP3 [dBm] | $\mathcal{N}_3$ [dBm/Hz] | $SFDR_2$ [dB·Hz$^{1/2}$] | $SFDR_3$ [dB·Hz$^{2/3}$] |
|---|---|---|---|---|---|---|---|---|
| InP EAM, $V_b =$ 1 V | 2 | −13.5 | 4.14 | 36.3 | 2.5 | −157.7 | 94.2 | 106.8 |
| $LiNbO_3$ MZM, push-pull | 2 | −18.9 | 2.74 | — | 1.8 | −160.4 | — | 108.1 |
| Si depletion MZM | 2 | −30.4 | 1.37 | 35.8 | −3.9 | −164.3 | 98.2 | 107.0 |
| DP-MZM, SSB | 2 | −33.0 | 1.09 | — | −9.3 | −165.5 | — | 104.1 |
| DML, $3.5I_{th}$ (Part I) | 2 | −13.9 | 13.75 | 23.5 | 24.7 | −148.7 | 82.8 | 115.6 |
| InP EAM, $V_b =$ 1 V | 10 | −13.8 | 4.14 | 36.3 | 2.5 | −144.2 | 95.1 | 97.8 |
| $LiNbO_3$ MZM, push-pull | 10 | −19.8 | 2.74 | — | 1.8 | −147.8 | — | 99.7 |
| Si depletion MZM | 10 | −31.1 | 1.37 | 35.8 | −3.9 | −153.6 | 98.9 | 99.8 |
| DP-MZM, SSB | 10 | −33.9 | 1.09 | — | −9.3 | −155.6 | — | 97.5 |
| DML, $3.5I_{th}$ (Part I) | 10 | −12.0 | 13.75 | 11.9 | 13.4 | −133.8 | 78.0 | 98.1 |

The dependence on the operating point is shown in Fig. 7. For the EAM [Fig. 7(a)], $SFDR_2$ peaks at 115.4 dB·Hz$^{1/2}$ at $V_b = 0.99$ V and $SFDR_3$ at 122.0 dB·Hz$^{2/3}$ at 1.71 V, at the null biases predicted by Eq. (19). The peaks are very narrow: a 40 mV shift from the IMD2 null reduces $SFDR_2$ to 88 dB·Hz$^{1/2}$. At the IMD3 null the exact $SFDR_3$ is limited by fifth-order distortion, whereas the small-signal value (dotted) diverges. The peak values are resolved to the 10 mV step of the bias sweep. For the push-pull MZM [Fig. 7(b)], the ideal cancellation of IMD2 at quadrature is fragile: a bias error of 1° limits $SFDR_2$ to 94.6–94.7 dB·Hz$^{1/2}$, 5° to 87–88 dB·Hz$^{1/2}$ and 20° to 80–83 dB·Hz$^{1/2}$, whereas $SFDR_3$ changes by less than 3 dB over ±20°. For the Si MZM [Fig. 7(c)], $SFDR_2$ is 98.2 dB·Hz$^{1/2}$ at quadrature and reaches 122.8 dB·Hz$^{1/2}$ at a heater offset of $-0.40°$, where the absorption-induced IMD2 is cancelled. This null is extremely narrow: $SFDR_2$ is within 10 dB of its maximum only over 0.02°, which would require active bias control. Because the IMD2 at quadrature scales with $r$, $SFDR_2$ is a direct probe of the amplitude-to-phase ratio of the junction. Finally, $SFDR_3$ decreases towards the laser relaxation frequency for all links, including the external ones [Fig. 7(d)], and reaches a common minimum of 96.3–98.6 dB·Hz$^{2/3}$ near 8.5 GHz. This minimum is set by the RIN peak of the CW laser, which the modulators transfer to the detected signal. Near the minimum the MZM and Si links exceed the DML by up to about 1.7 dB. At lower and higher frequencies the intrinsic DML model is equal or superior, with the caveat discussed above.

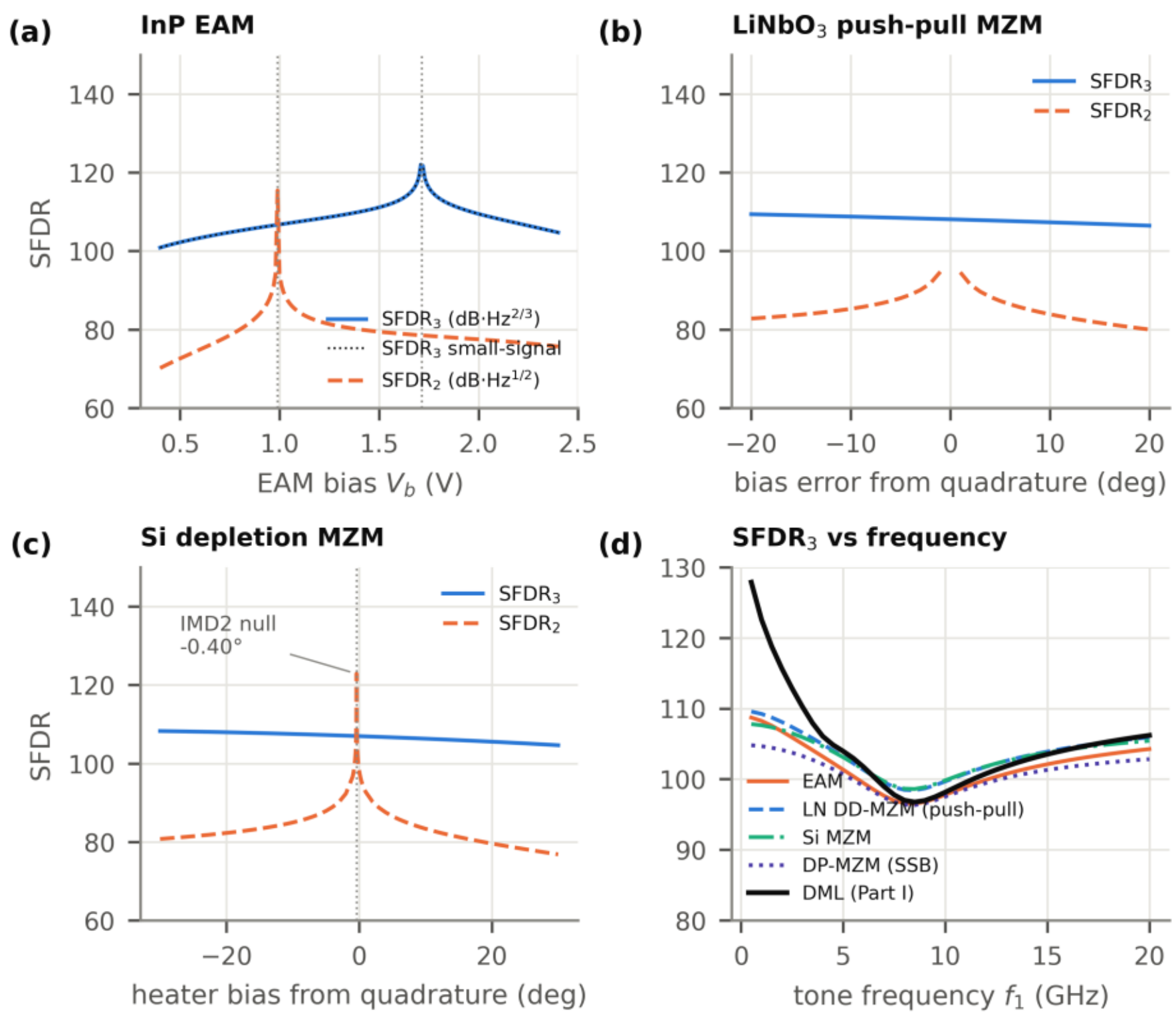

**Fig. 7.** SFDR in 1 Hz versus operating point at $f_1 = 2$ GHz: (a) InP EAM versus bias voltage (exact $SFDR_3$, solid; small-signal $SFDR_3$, dotted; exact $SFDR_2$, dashed), with the null biases of Eq. (19) marked; (b) push-pull $LiNbO_3$ MZM versus bias error from quadrature; (c) Si depletion MZM versus heater bias (0.01° steps within ±1°). (d) $SFDR_3$ versus tone frequency for the four external modulators and the DML of Part I, all driven by or fed from the same laser at $3.5I_{th}$.

## 8. Discussion

The framework of Part I extends to external modulation with one essential simplification. Because an external modulator is memoryless after its electrode, its chirp is entirely described by the chirp map $\alpha_m(v)$, with no adiabatic component, and its Volterra kernels factorize into Taylor coefficients and electrode responses. The chirp map is a natural generalization of the modulator chirp parameter [6]. It contains the voltage dependence that matters for large-signal digital drive, which a single chirp parameter measured at the operating point cannot capture [7]. Equation (13) shows that a time–frequency measurement of the field under ordinary data modulation returns the map over the voltage range swept by the transitions. For the Si modulator, the value of the map at quadrature, $-r$, is a direct measure of the Soref–Bennett amplitude-to-phase ratio of the junction.

In the time–frequency plane the four modulators have distinct signatures. The EAM produces transient strokes whose sign follows the bias, with no level-dependent offset. Push-pull MZMs produce distributions symmetric about the carrier, and single-drive MZMs produce strokes at every transition, proportional to the drive slew rate. The silicon modulator shows only a weak residual chirp, and complex modulation with the DP-MZM produces phase-jump spikes at the zero crossings of the field. These signatures are the time–frequency counterpart of the well-known dispersion tolerance of each format. Together with the fiber shear of the Wigner distribution mentioned in Part I, they provide a route to a unified time–frequency description of dispersive propagation.

For analog links, the closed-form intercepts and bias sensitivities reproduce the classical results for MZMs [3,18] and add simple expressions for the EAM null biases and for the intrinsic second-order distortion of depletion modulators [12,13]. A result that follows from combining the two papers is that the laser RIN transfers unchanged to externally modulated links. With a low-power DFB source biased at a few times threshold, as here, the RIN peak sets a common SFDR minimum near the laser relaxation frequency. This is a property of the chosen source, not of external modulation: analog external links normally use high-power, low-RIN lasers whose relaxation frequency lies far above the band of interest. With the low photocurrents that result from insertion loss and quadrature bias, the external links considered here are shot- and RIN-limited, and they do not reach the intrinsic upper bound of the DML model. Their practical advantages are the absence of chirp or controllable chirp, bandwidths beyond the laser relaxation frequency, the absence of the static nonlinearities of a directly modulated laser and, for SSB, immunity to dispersion-induced fading [9], which is outside the back-to-back analysis presented here. Phase-to-intensity noise conversion is also absent in the balanced MZMs and in the back-to-back links considered.

The limitations are the following. The models are phenomenological: the EAM transfer and chirp laws, the silicon depletion law with a constant amplitude-to-phase ratio (which directly sets its chirp at the bias point and its second-order distortion) and the traveling-wave electrode response neglect saturation, photocurrent effects and thermal drift in the EAM, and impedance mismatch and electrode dispersion in the MZMs. The parameters are representative, not measured. The links are back-to-back, and the noise model includes only laser RIN, shot and thermal noise.

## 9. Conclusion

We have extended the time–frequency framework of Part I to the external modulation of a CW semiconductor laser by InP electroabsorption, $LiNbO_3$ dual-drive and dual-parallel, and silicon depletion modulators. For any memoryless modulator the chirp is purely transient, $\Delta\nu = \alpha_m(v)(4\pi)^{-1}d\ln P/dt$, with a chirp map $\alpha_m = \mathrm{Im}(M'/M)/\mathrm{Re}(M'/M)$. We derived the map in closed form for each device, including $\alpha_m = -r$ for the silicon modulator at quadrature. The centroid-to-instantaneous-bandwidth ratio of a Cohen-class distribution recovers the map from a single data record within 0.08. Jacobi–Anger line expansions of the dual-drive and dual-parallel MZMs agree with numerical spectra to machine precision. Factorized Volterra kernels give closed-form intercepts that match the numerical ones to within 0.01 dB, as well as the EAM linearization biases, $V_0/\sqrt{2}$ and $\sqrt{3/2}\,V_0$. Fed by the DFB laser of Part I, the external links reach $SFDR_3$ of 104–108 dB·$Hz^{2/3}$ at 2 GHz, below the intrinsic upper bound of the DML model, and their dynamic range degrades near the relaxation frequency of the source because the modulators transfer its RIN. A small heater offset cancels the absorption-induced second-order distortion of the silicon modulator, but only within a 0.02° window.

**Funding.** [To be completed by the author.]

**Acknowledgment.** [Optional; to be completed by the author.]

**Disclosures.** The author declares no conflicts of interest.

**Data availability.** Data and simulation code underlying the results presented in this paper are not publicly available at this time but may be obtained from the author upon reasonable request.